\documentclass[%
 reprint,
superscriptaddress,
amsmath,amssymb,
aps,
prb,
floatfix,
]{revtex4-2}
\usepackage{float}

\usepackage{graphicx}
\usepackage{dcolumn}
\usepackage{bm}
\usepackage[dvipsnames]{xcolor}
\usepackage[colorlinks=true]{hyperref}
\hypersetup{
    colorlinks=true,
    linkcolor=blue,
    citecolor=blue,
    urlcolor=blue
} 

\begin{document}
\preprint{APS/123-QED}
\title{ Low-Frequency Noise and Resistive Switching in $\beta$-Na$_{0.33}$V$_2$O$_5$}
\author{Nitin Kumar}
\affiliation{%
Department of Physics, University at Buffalo-SUNY, 239 Fronczak  Hall, Buffalo, NY 14260 US}%

\author{Nicholas Jerla}
\affiliation{Department of Physics, University at Buffalo-SUNY, 239 Fronczak  Hall, Buffalo, NY 14260 US}%

\author{John Ponis}
\affiliation{
 Department of Chemistry, Texas A\&M University, College Station, TX 77843, USA
}%

\author{Sarbajit Banerjee}%
\affiliation{
 Laboratory for Inorganic Chemistry, Department of Chemistry and Applied Biosciences, ETH Zurich, Vladimir-Prelog-Weg 2, CH-8093 Zürich, Switzerland
}%
\affiliation{Laboratory for Battery Science, PSI Center for Energy and Environmental Sciences, Paul Scherrer Institute, Forschungsstrasse 111, CH-5232 Villigen PSI, Switzerland}
\affiliation{Department of Material Science and Engineering, Texas A\&M University, College Station, TX 77843, USA
}%

\author{G. Sambandamurthy}
\email{sg82@buffalo.edu}
\affiliation{Department of Physics, University at Buffalo-SUNY, 239 Fronczak  Hall, Buffalo, NY 14260 US}

\begin{abstract}
The interplay between charge ordering and its manifestation in macroscopic electrical transport in low-dimensional materials is crucial for understanding resistive switching mechanisms. In this study, we investigate the electronic transport and switching behavior of single-crystalline $\beta$-Na$_{0.33}$V$_2$O$_5$, focusing on low-frequency resistance noise dynamics of charge-order-driven resistive switching. Using electrical transport, low frequency noise spectroscopy, and X-ray diffraction, we probe electron dynamics across the Na-ion- ordering (IO) and charge-ordering (CO) transitions. Near room temperature, the weak temperature dependence of the noise spectral density points to a dominance of nearest-neighbor polaron hopping. Below IO transition temperature (\( T_{IO} \sim 240 \, \text{K} \)), structural analysis reveals that Na-ions adopt a zig-zag occupancy pattern, breaking the two-fold rotational symmetry and influencing the electronic ground state. Subsequently, a sharp drop in resistance noise below the CO transition temperature (\( T_{CO} \sim 125 \, \text{K} \)) indicates the emergence of correlated electron behavior. Furthermore, application of sufficient electric field leads to the destabilization of the CO state, and a transition to a high-conducting state. The material exhibits distinct resistive switching between 35~K and 110~K, with a resistance change spanning two orders of magnitude, primarily driven by electronic mechanisms rather than Joule heating. These findings provide new insights into charge-order-induced switching and electronic correlations in quasi-one-dimensional systems, with potential applications in cryogenic memory and neuromorphic computing devices owing to the low noise levels in their stable resistive states.
\end{abstract}

\maketitle

\section{\label{sec:level1}Introduction}

Low-dimensional systems with strong electronic correlations exhibit a variety of unconventional phenomena, including charge ordering, polaronic behavior, and metal-insulator transitions.\cite{alder2018, Moser2013,sood2021} Electronic phase transitions have significant effects on transport properties and are sensitive to subtle modulation of external parameters   such as temperature, pressure, or electric field.\cite{Lee2018,vale2019,dagotto2005} The competition between localization, electronic correlations, and external driving forces needs to be clarified to disentangle the mechanisms that govern conduction in low-dimensional systems.\cite{dagotto2005,Alam2023} In the past decade, correlated oxides and mixed-valence compounds have attracted wide attention due to their promise to exhibit tunable resistive states and electric field-driven switching behavior, \cite{Nitin2025,sawa2008,waser2009,janod2015,kaushik2025} making them as attractive candidates for fundamental studies of non-equilibrium transport and as next-generation electronic devices that operate across a range of ambient environments.\cite{Hutter_2011,Jo2010,lanyon2008,Pathak2018,Preskill_2018} However, a clear understanding of how microscopic electronic phenomena, such as charge ordering, Mott transitions, and polaron formation, influence macroscopic transport properties and resistive switching behavior remains elusive. This gap is especially pronounced in quasi-one-dimensional correlated systems, where the interplay of charge, lattice, and spin disorder leads to complex and emergent electronic responses.\cite{Geremew2019_ACSNano}

The quasi-one-dimensional $\beta$ ($\beta'$)-A$_x$V$_2$O$_5$ (A = Li, Na, Ca, Pb, Cu, etc.) vanadium oxide bronzes represent a class of materials exhibiting resistance changes spanning several orders of magnitude in response to temperature and electric field.\cite{Marley2014,PARIJA20201166,Marley2013,Andrew2019} In these bronzes, the electrical resistance can be tuned by the choice of intercalant, giving rise to a wide range of electrical and magnetic behaviors within this material family.\cite{Ma2006} The crystals feature a distinctive layered tunnel structure, characterized by high anisotropy, enabling novel phenomena such as a high-temperature metallic state observed when specific stoichiometric ratios of inserted cations, typically $x = 0.33$ or $0.65$ for Cu insertion in the $\beta'$-phase, are incorporated into the tunnel.\cite{Okazaki2004} For Na inserton at $x = 0.33$, the average structure contains one Na-ion randomly distributed within every pair of neighboring $\beta$-sites, resulting in a half-occupied Na-ion sublattice stabilized by local zig-zag ordering.\cite{Krogstad2020} This arrangement induces a long-range periodic potential parallel to the $b$-axis throughout the tunnel structure, giving rise to a quasi-one-dimensional metallic state at room temperature.\cite{Ma2006,Okazaki2004}  However, this metallic state is highly sensitive to off-stoichiometry and crystal defects, which disrupt the Na-induced long-range potential and pattern of electron and spin localization along the host $\zeta$-V$_2$O$_5$ framework, which significantly affects the material’s electrical properties.
\vspace{-0.1cm}
\begin{figure}
    \centering
    \includegraphics[width=0.7\linewidth]{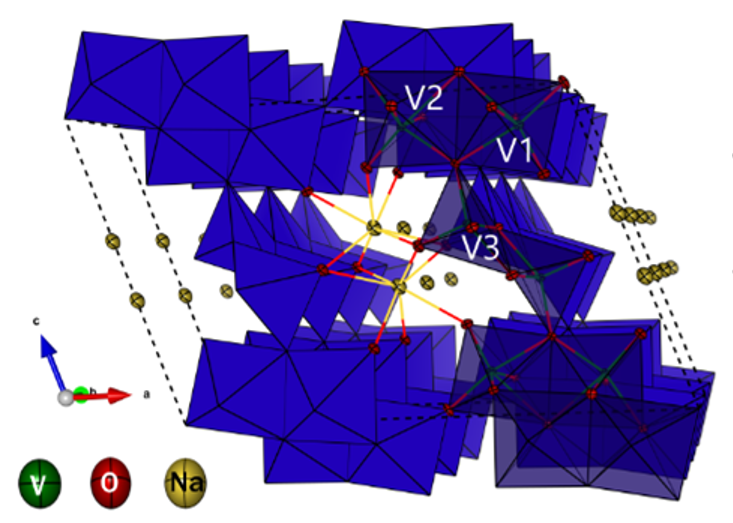} 
    \caption{ Structure of $\beta$-Na$_{0.33}$V$_2$O$_5$, highlighting the arrangement of Na-sites within a tunnel structure and V-O coordination geometry.}
    \label{fig:structure}
\end{figure}
Additionally, these vanadium bronzes are mixed-valence compounds, with V$^{4+}$ (3d$^1$) and V$^{5+}$ (3d$^0$) distributed between the V1 and V2 sites. These sites exhibit octahedral coordination with the adjacent oxygen atoms. The (V1)O$_6$ octahedra form a zig-zag double-chain structure, whereas the (V2)O$_6$ octahedra create a two-legged ladder structure, aligned parallel to the $b$-axis. The non-magnetic V3 site, which has a formal valence of V$^{5+}$, forms zig-zagging square pyramidal coordination, also parallel to the $b$-axis. This unique arrangement leads to the tunnel structure observed in these materials, where the $\beta$-inserted-ion resides within the $\beta$-sites in the tunnels, as illustrated
in Fig. \ref{fig:structure}.\cite{Ma2006,Markina2004,Itoh2006,Okazaki2004,van2002,Presura2003} Notably, the $\beta$-sites cannot have both nearest-neighbor positions occupied simultaneously, because of cationic repulsions. The $\beta^{\prime}$-sites, which are always farther apart than the $\beta$-sites, are typically occupied when cation stoichiometry exceeds 0.33 as in the case of Li- and Cu-inserted $\beta^{\prime}$-A$_x$V$_2$O$_5$ compounds.

To explore correlated charge phenomena, low-frequency resistance noise spectroscopy has proven to be a powerful probe for detecting dynamic fluctuations and phase transitions in complex materials.\cite{Geremew2019_ACSNano,singh2016,Alsaqqa2017} Unlike conventional transport measurements, resistance noise reveals hidden information about carrier dynamics, disorder, and correlation effects. In particular, a sharp drop in noise magnitude is recognized as a signature of collective or glassy charge dynamics in correlated systems.\cite{Samanta2012_PRB} In parallel, charge ordering in low-dimensional oxides has been linked to field-induced resistive switching, highlighting the role of electronic phase transitions in enabling non-volatile conductance states.\cite{Vaskivskyi2016_NatCommun} These insights motivate a deeper investigation of switching mechanisms in charge-ordered materials such as $\beta$-Na$_{0.33}$V$_2$O$_5$.

Upon cooling from room temperature, $\beta$-Na$_{0.33}$V$_2$O$_5$ undergoes a sequence of phase transitions. Initially, the Na-ions, which occupy one of two distinct crystallographic positions within the tunnel network, \cite{Markina2004} begin to structurally order at $T_\text{IO} \sim 230$~K, \cite{Yamada1999, Markina2004} forming a zig-zag pattern that alternates between $\beta$-sites along the $b$-axis. As the temperature is further reduced below $T_\text{IO}$, the Na-ion ordering becomes complete, eventually leading to charge redistribution primarily onto the V1 zig-zag chains.\cite{Presura2003, Natanzon2020} Upon further cooling, a charge ordering (CO) transition occurs at $T_\text{CO} \sim 130$~K, \cite{van2002, Yamada1999} which is accompanied by the localization of carriers. Optical spectroscopy suggests that these charge carriers may form small polarons, underscoring the significant role of electron–phonon interactions in stabilizing the CO ground state.\cite{Kuntscher2005} At $T_\text{N} \sim 22$~K, the system undergoes spin ordering, forming a canted antiferromagnetic structure.\cite{Yamauchi2002} Finally, under hydrostatic pressures exceeding 8 GPa, a superconducting phase emerges below 10 K.\cite{Okazaki2004,Yamauchi2002,Itoh2006,Sirbu2006} The coexistence of strong correlations, reduced dimensionality, and tunable ground states render this system a fertile ground for fundamental studies and for exploring its potential as a platform for memory and computing applications, motivating further investigation of its complex phase behavior.

In this work, we investigate the resistive switching behavior and low-frequency noise dynamics of $\beta$-Na$_{0.33}$V$_2$O$_5$, focusing on the interplay between charge ordering and conduction mechanisms at low temperatures. Using electrical transport measurements in conjunction with low-frequency noise spectroscopy, we examine the evolution of charge transport across the CO transition. Our analysis reveals a crossover from thermally activated polaronic conduction at higher temperatures to a correlated electronic state below the CO transition. Notably, an abrupt suppression of noise magnitude near $T_{\mathrm{CO}}$, indicates the onset of collective charge dynamics in the ordered phase. Complementary structural evidence from temperature-dependent X-ray diffraction reveals the development of satellite reflections consistent with the formation of a modulated superstructure, corroborating the electronic transition inferred from transport and noise signatures. Furthermore, the insulating CO state can be destabilized by an electric field, inducing a transition to a high-conducting state with resistive switching characteristics. The voltage-driven noise spectra reveal a steepening of the $1/f^\alpha$ behavior toward $\alpha \sim 2$ at the onset of switching, indicative of slow, cooperative dynamics associated with the breakdown of charge order. These findings provide new insights into charge localization, correlated transport, and field-driven phase transitions in quasi-one-dimensional systems, contributing to a deeper understanding of resistive switching phenomena in correlated materials.

\section{Experimental Methods}

Single crystals of $\beta$-Na$_{0.33}$V$_2$O$_5$ were prepared by a solid-state powder synthesis and subsequent melt-growth, as described in .\cite{Parize1989} Na$_2$C$_2$O$_4$ (Sigma-Aldrich) and V$_2$O$_5$ (Sigma-Aldrich) at a 0.18:1 molar ratio (a small stoichiometric excess of Na was used to maximize Na-site filling) were ground together thoroughly in a mortar and pestle and loaded into an alumina combustion boat. The mixture was placed in an ambient-temperature tube furnace, held under vacuum, and heated to 550°C for 16 h. After cooling, the sample was removed, ground, and heated as previously. The identity of the resulting blue-black powder was confirmed by powder X-ray diffraction. To prepare single crystals, the powder product was sealed in an evacuated amorphous silica ampoule, melted at 810°C in a muffle furnace, cooled to 550°C at 2°C/h to crystallize, and cooled to ambient temperature under furnace momentum. Several regular, lustrous black acicular crystals were harvested for measurement.
\begin{figure*}
    \centering
    \includegraphics[trim=15 10 12 5, clip, width=\linewidth]{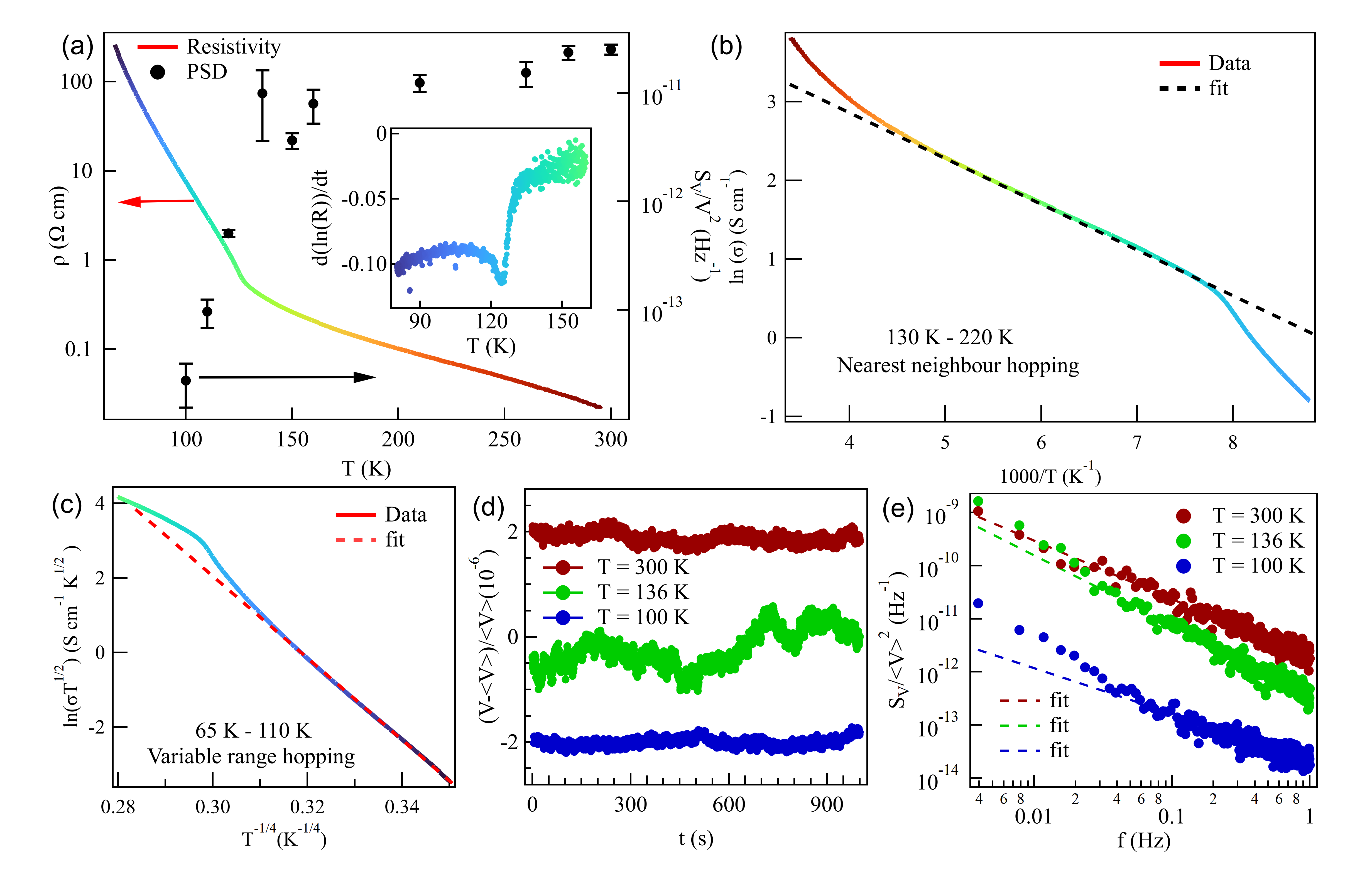}
    \caption{(a) Temperature dependence of resistivity and noise. Error bars represent the standard deviation from three different measurements. (Inset: The derivative of resistance with respect to temperature, highlighting a transition around 125 K). (b) Fitting of the resistivity data to a small polaron nearest-neighbor hopping (NNH) model in the high-temperature regime. (c) Fitting of the resistivity data to a variable-range hopping (VRH) model in the low-temperature regime. (d) Time series of residual voltage fluctuations from the mean at three distinct temperatures (The plots at 300 K and 100 K are shifted upward and downward, respectively, for clarity). (e) Power spectral density of the residual voltage fluctuations derived from (d), illustrating their frequency-dependent ($1/f^\alpha$) noise.}
    \label{fig:resistivity_noise}
\end{figure*}
Diffraction data were collected using a Bruker D8 Quest diffractometer with the APEX4 software suite and Mo K$\alpha$ ($\lambda = 0.71073$ Å) X-ray tube source. Temperature control was achieved using an Oxford Instruments N$_2$ cryostream. Data for structure solutions were collected at each temperature using $\omega$ and $\phi$ scans. Frames were integrated, scaled, and merged using APEX4. Structures were solved by intrinsic phasing using SHELXT and refined on F$^2$ with SHELXL, implemented in Olex2.\cite{Sheldrick2008, Dolomanov2009}

X-ray data over the interval from 100 K-170 K were collected by first cooling the crystal from ambient temperature to 100 K at 6 K/min and collecting data for the 100 K structure solution. The temperature was then increased at 6K/min to 170 K, pausing at temperature intervals to collect an $\omega$ scan for supercell reflection intensity quantitation. Once at 170 K, data were collected for structure solution, and the crystal was cooled back to 100 K with incremental $\omega$ scans. Data over the interval 170 K-290 K were collected for the same crystal in a subsequent experiment. Data for structure solution were collected at 290 K, and the crystal was cooled to 170 K and then heated to 290 K with incremental $\omega$ scans as before. Frames collected at temperature intervals were integrated using APEX4, with sequential unit cell parameter refinement. Reflections with signal-to-noise ratio $I/\sigma > 3$ were binned according to modulation vector, summed, and normalized to the sum of all reflection intensities.

For electrical transport measurements, crystals were secured to a glass substrate using GE varnish with electrodes made by painting conducting Ag paste onto the device. Cu wires were used to connect the electrodes to the electrical pins of the sample probe. Transport measurements were performed in a closed-cycle, superconducting magnet system from LakeShore/Janis using LakeShore 336 temperature controllers with 10 mK stability during all measurements. Current-voltage characteristics were measured using a Keithley 2450 SMU and noise measurements were measured with an SRS 7265 DPS lock-in amplifier. Low noise decade resistors were used as the ballast and variable resistors in our bridge circuit for noise measurements.

Low-frequency resistance noise was measured using a Wheatstone bridge method with our sample as one of the resistors in the arms of the bridge. Data were sampled through an ADC at 256 Hz from the lock-in with an internal low-pass filter whose roll-off time constant was set to 20 ms over a time of 16 min and decimated down to an 8 Hz sample rate to reduce aliasing and attenuate any unwanted signal above 8 Hz. An oscillation frequency of 283.117 Hz was supplied to avoid power line harmonics and be attenuated when the time series is processed. A series of digital processing techniques were then used to convert the raw time series into a power spectral density (PSD, $S_v(f)$) as a function of frequency.

\section{RESULTS AND DISCUSSION}

\begin{figure*}
    \centering
    \includegraphics[width=\linewidth]{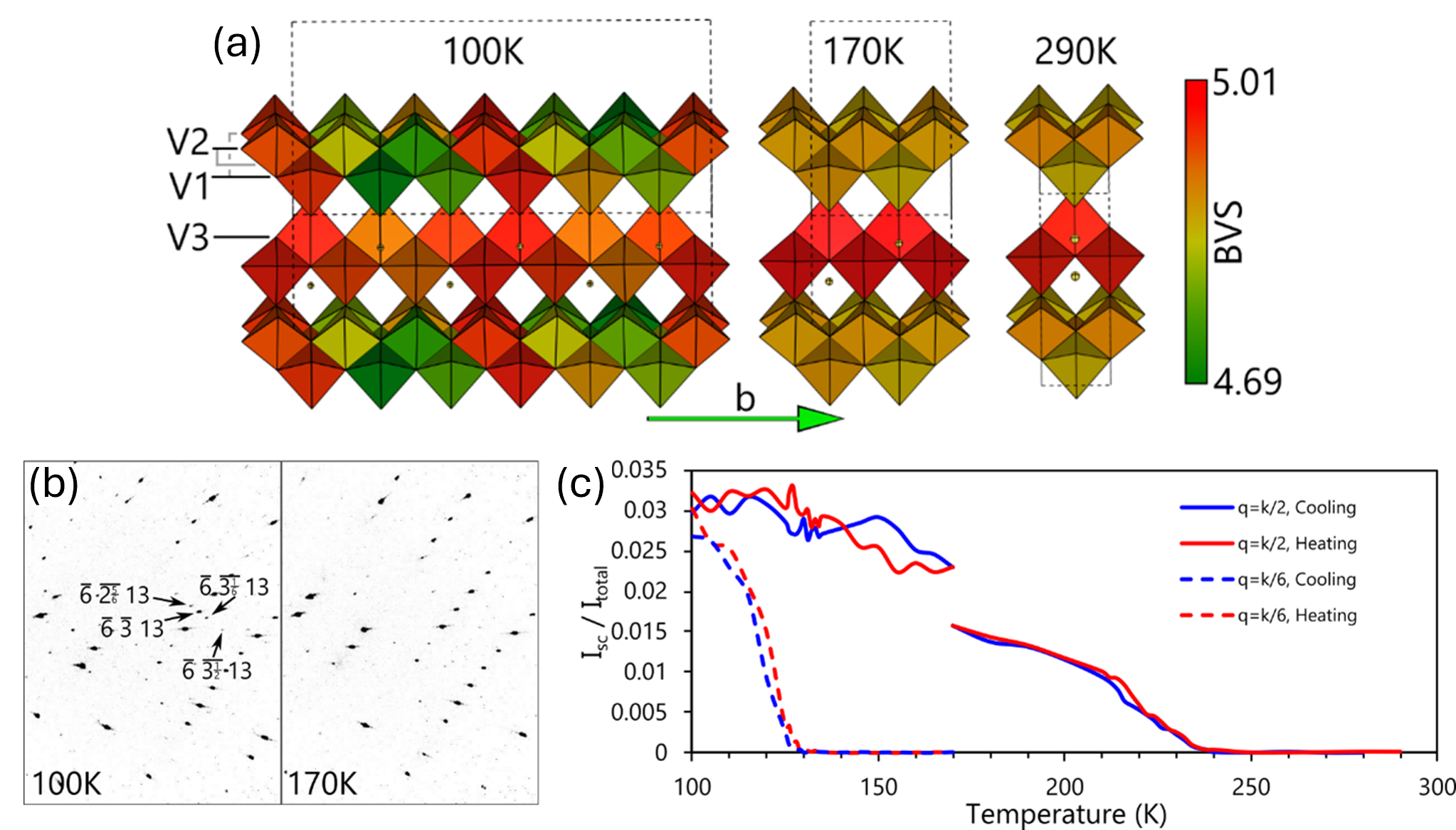} 
    \caption{Temperature-dependent structure transitions in $\beta$-Na$_{0.33}$V$_2$O$_5$. (a) Structure solutions at 100 K, 170 K, and 290 K, with VO$_x$ polyhedra colored according to bond valence sum. Zig-zag ordering of Na-ions and periodic distortions to V–O local structure contribute respectively to 2- and 6-fold cumulative expansions of the unit cell (dashed lines) along $b$. (b) Detector frame images collected at 100 K and 170 K, with several modulation satellites indicated by arrows. (c) Relative modulation satellite intensity vs. temperature. The discontinuity at 170 K results from crystal orientation differences between separate measurements.}
    \label{fig:structure_transitions}
\end{figure*}
The temperature dependence of resistivity along the \textit{b}-axis, measured from 300 K to 50 K as shown in Fig. \ref{fig:resistivity_noise}(a), reveals a nonlinear change upon cooling, attributed to a semiconducting-to-insulating charge-ordering (CO) transition occurring at \( T_{CO} \sim 125 \, \text{K} \) (as shown in the inset of Fig. \ref{fig:resistivity_noise}(a)).

Above \( T_{CO} \), the data were fit to the nearest-neighbor hopping (NNH) model for small polarons by plotting \( \ln(\rho/T) \) versus \( T^{-1} \), as shown in Fig. \ref{fig:resistivity_noise}(b), following Eq. \ref{eq:NNH}.\cite{Natanzon2020} The extracted activation energy is approximately 63 meV, in agreement with previously reported values for $\beta$-Na$_{0.33}$V$_2$O$_5$.\cite{Shukla2022, Sirbu2006} This energy represents the thermal barrier for polaron hopping and reflects the degree of carrier localization, making it a critical parameter for understanding the transport mechanism in the high-temperature regime.

\begin{equation}
\rho(T) = \rho_0 T \exp\left( \frac{E_a}{k_B T} \right)
\label{eq:NNH}
\end{equation}
However, as shown in Fig. \ref{fig:resistivity_noise}(b), the NNH fit is valid only in the temperature range from 130 K to 220 K, suggesting the presence of another transition near this range, which is discussed later. The variable-range hopping (VRH) conduction mechanism below \( T_{CO} \) is analyzed by plotting \( \ln(\rho/T^{1/2}) \) against \( T^{-1/4} \), following the temperature dependence described by Eq.~\ref{eq:VRH}, \cite{Shukla2022, RN94} as depicted in Fig. \ref{fig:resistivity_noise}(c).

\begin{equation}
\rho(T) = \rho_0 \exp\left[ \left( \frac{T_0}{T} \right)^{1/4} \right]
\label{eq:VRH}
\end{equation}
This analysis yields Mott’s characteristic temperature \( T_0 \), which is found to be \( 1.88 \times 10^8 \, \text{K} \). Based on this value, the density of states at the Fermi level \( N(E_f) \), the mean hopping distance \( R \), and the mean hopping energy \( W \) were calculated. Assuming that the small polaron is localized to a single vanadium site, the localization length was approximated as \( \alpha^{-1} \sim 1 \, \text{Å} \).\cite{RN94} These calculations yielded \( N(E_f) = 1.16 \times 10^{21} \, \text{eV}^{-1} \text{cm}^{-3} \), \( R = 1.27 \, \text{nm} \), and \( W = 100.47 \, \text{meV} \). Additionally, for the VRH model to remain valid, the conditions \( \alpha R > 1 \) and \( W > k_B T \) were verified for all temperatures below \( T_{CO} \), as outlined in Eq.~\ref{eq:VRH_limits}.\cite{Shukla2022,DKPAUL1973}

\begin{subequations}\label{eq:5}
\begin{align}
k_B T_0 &= \frac{18 \alpha^3}{N(E_f)} \label{eq:5a} \\
W &= \frac{3}{4 \pi R^3 N(E_f)} \label{eq:5b} \\
R &= \left( \frac{9}{8 \pi \alpha k_B T N(E_f)} \right)^{1/4} \label{eq:5c}
\end{align}
\label{eq:VRH_limits}
\end{subequations}

To probe the carrier dynamics across the transition, low-frequency noise measurements were performed over the temperature range of 100 K to 300 K. Fig. \ref{fig:resistivity_noise}(d) displays representative residual current fluctuations measured above, near, and below the transition, with notably higher fluctuations observed in the vicinity of \( T_{\text{CO}} \). The PSD of these fluctuations are shown in Fig. \ref{fig:resistivity_noise}(e), exhibiting a clear inverse frequency dependence. Fig. \ref{fig:resistivity_noise}(a) shows the PSD magnitude at 0.1 Hz as a function of \textit{T}, revealing a weak temperature dependence above \( T_{\text{CO}} \) and a sharp suppression of noise as the system enters the charge-ordered state below \( T_{\text{CO}} \). This weak temperature dependence at high-temperatures  is characteristic of hopping conduction in systems \cite{Shklovskii2003, RN97} where transport is dominated by thermally activated hopping between nearest-neighbor localized states. In this regime, the resistivity becomes nearly temperature-independent, and the conduction pathways stabilize, leading to minimal variation in noise magnitude. The strong suppression of noise below \( T_{\text{CO}} \), in contrast, reflects the onset of enhanced carrier localization and collective electronic correlations-providing noise-based evidence for the transport crossover from NNH to VRH, as suggested by the resistivity analysis.

Similar to optical measurements showing a gradual ordering of the Na-ions, \cite{Presura2003} further insights into the temperature dependence of the IO and CO can be probed through X-ray diffraction intensities, as shown in Fig. \ref{fig:structure_transitions} (c). These intensities reveal a gradual increase in CO due to enhanced electron localization as the system is further cooled below \( T_{CO} \). This progressive charge localization at lower temperatures underpins for the observed pressure-temperature (\( P-T \)) diagrams in many other organic and heavy-fermion systems exhibiting pressure-induced superconductivity.

Recent reports have also discussed an additional transition corresponding to Na-ion ordering (IO) at \( T_{IO} \sim 240 \, \text{K} \).\cite{Obermeier2002,Heinrich2004} Optical measurements have shown that this transition occurs gradually from \( T_{IO} \) to \( T_{CO} \), where charge localization occurs along the zig-zag chains of vanadium atoms spanning the $b$-axis, centered on the \( V_1 \) vanadium atom.\cite{Yamada1999} Single-crystal X-ray diffraction measurements reveal the atomic structure modulations exhibited by the low-temperature insulating state. $\beta$-Na$_{0.33}$V$_2$O$_5$ exhibits the C2/m space group at room temperature, shown in Fig. \ref{fig:structure_transitions} (c). (V1)O$_6$ and (V2)O$_6$ octahedra form V$_4$O$_{16}$ clusters, which edge-share in chains along $b$, connected along $a$ by shared V1-O2-V1 corners and bridged along $c$ by zig-zag chains of edge-shared (V3)O$_5$ square pyramids. Na-ions occupy paired sites within tunnel-shaped interstices extending along $b$. The proximity of these site pairs (1.953 \AA) limits them to half-occupancy, corresponding to a nominal Na$_{1/3}$V$_2$O$_5$ composition.

Upon cooling below $T_{\text{Na}} = 240$ K, Na-ions adopt a zig-zag occupancy pattern in the average structure, breaking the 2-fold rotation symmetry related to the half-occupied sites, generating the P2$_1$/c space group and forming an ($a$, $2b$, $c$) supercell (Fig. \ref{fig:structure_transitions} (a)). As a result, satellite reflections appear at $\mathbf{q} = (h, k/2, l)$, visible in the right panel of Fig. \ref{fig:structure_transitions} (b). A recent three-dimensional difference pair distribution function (3D-$\Delta$PDF) study of diffuse scattering intensity (visible as streaking in Fig. \ref{fig:structure_transitions} (b)) demonstrates that this zig-zag ordering is present in individual tunnels above $T_{\text{IO}}$, but that translation symmetry with adjacent unit cells is frustrated at high temperature by the degeneracy of ordering regimes in intervening tunnels.\cite{Krogstad2020} The gradual increase in relative intensity of $\mathbf{q} = (h, k/2, l)$ reflections shown in Fig. \ref{fig:structure_transitions} (c) is consistent with continued growth of ordered domains as temperature decreases.
\begin{figure}
    \centering
    \includegraphics[trim=24 16 14 14, clip,width=0.9\linewidth]{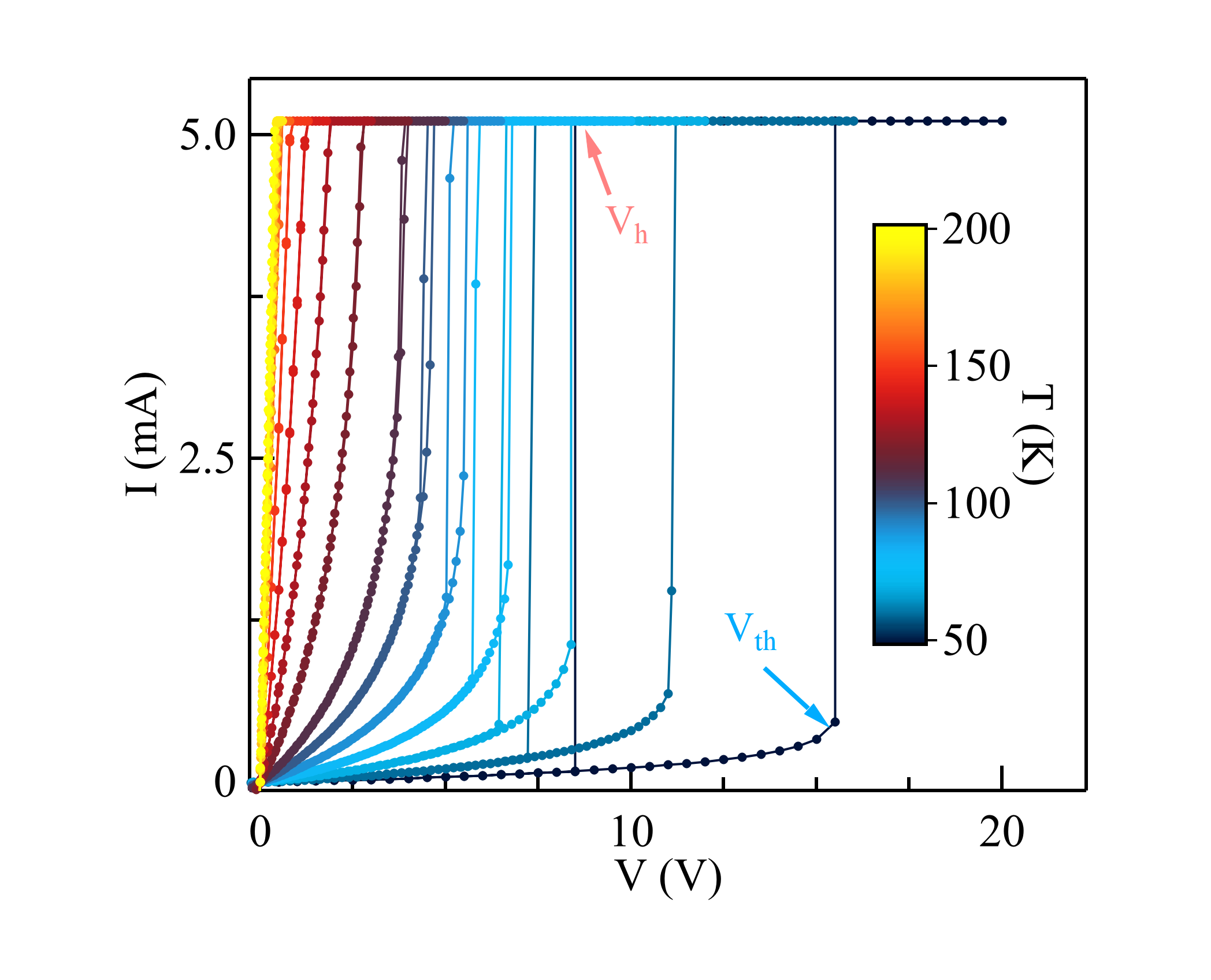} 
    \caption{Current-voltage characteristics, measured at various temperatures from \textit{T} = 50 K to 200 K, display abrupt resistance switching behavior. A current limit of 5 mA was set to prevent permanent damage to the device. \(V_{\text{th}}\) is the threshold voltage and \(V_{\text{h}}\) is hold voltage.}
    \label{fig:IV}
\end{figure}
Below $T_{\text{CO}} = 125$ K, the material adopts an ($a$, $6b$, $c$) supercell, as indicated by the presence of $\mathbf{q} = (h, k/6, l)$ satellite reflections at 100 K in Fig. \ref{fig:structure_transitions} (b). Vanadium bond valence sum \cite{Bard1999} calculations suggest the formation of an approximate $3b$ periodic CO in conjunction with the extant $2b$ Na ordering as the source of the modulation ( Fig. \ref{fig:structure_transitions} (a) ). CO has previously been invoked to describe the conductivity and structural transitions and others coincident at 125 K, \cite{van2002, Presura2003} and several charge distribution models have been proposed.\cite{Sirbu2006, Yamaura2002, Itoh2006} Notably, the charge distribution present is nearly identical to that supporting the antiferromagnetic ground state, as determined by single-crystal neutron diffraction below the Néel temperature.\cite{Nagai2005} This magnetic ordering likely coalesces from the CO pattern established at the onset of the semiconductor-insulator transition as presented above. Like the lower-order satellites, Fig. \ref{fig:structure_transitions} (c) shows that the reflections at $\mathbf{q} = (h, k/6, l)$ intensify with cooling, indicating deepening of charge localization.

In non-interacting systems, noise increases as the temperature decreases due to the narrowing of the density of states near the Fermi energy (\(E_f\)) at lower temperatures, which elevates the energy barriers for hopping. As a result, hopping events between localized states require longer timescales, leading to higher noise magnitudes. Essentially, the reduced availability of states for particles to hop causes higher noise levels at lower temperatures, characteristic of non-interacting systems.\cite{RN97,Burin2006} In contrast to characteristic non-interacting patterns, the observed noise behavior in single crystals of $\beta$-Na$_x$V$_2$O$_5$ exhibits pronounced differences. Indeed, the noise magnitude decreases below the characteristic transition temperature T$_{CO}$, a phenomenon that cannot be explained within the framework of non-interacting models. The observed anomaly is attributed to the onset of strong electron-electron correlations due to CO, which drive a collective reorganization of localized states under the influence of Coulomb interactions. In this correlated regime, these interactions suppress the independent hopping fluctuations characteristic of non-interacting systems, ultimately giving rise to a glassy state.\cite{Shtengel2003, RN100} As previously reported in optical studies, the observed mid-infrared absorption, combined with strong electron-phonon coupling and effective carrier density, suggests that small polaron transport plays a dominant role in $\beta$-Na$_{0.33}$V$_2$O$_5$.\cite{Presura2003} Hall measurements have revealed that the total number of charge carriers remains largely unaffected by temperature changes in the range of 189 K to 298 K, with the carrier density approximately matching the concentration of Na atoms in the crystal.\cite{Perlstein1968} This gives additional support to the polaronic nature of conduction, as each excess electron introduced into the lattice by the intercalated monovalent Na localizes on the \( V_1/V_2 \) vanadium sites, which exhibit a mixed valence of +4/+5.

The progressive strengthening of CO upon cooling, as evidenced by the increasing satellite intensities reflecting increased dimensions ordered domains, enhances the potential of the already localized charges, thereby further impeding charge hopping between sites. However, this CO state can be disrupted by applying sufficient external stimuli, such as voltage. To investigate this phenomenon, current-voltage (\(I\)-\(V\)) characteristics were measured, with the results presented in Fig. \ref{fig:IV}. As the applied voltage increases, the current initially rises nonlinearly until a threshold voltage (\(V_{\text{th}}\)) is reached, at which point a sudden increase in current is observed, indicating resistive switching resulting from breaking of charge ordering. Upon reducing the voltage below hold voltage (\(V_{\text{h}}\)), the sample reverts to an insulating state, resulting in a sudden drop in current. 
\begin{figure*}
    \centering
    \includegraphics[trim=20 8 17 10, clip, width=\linewidth]{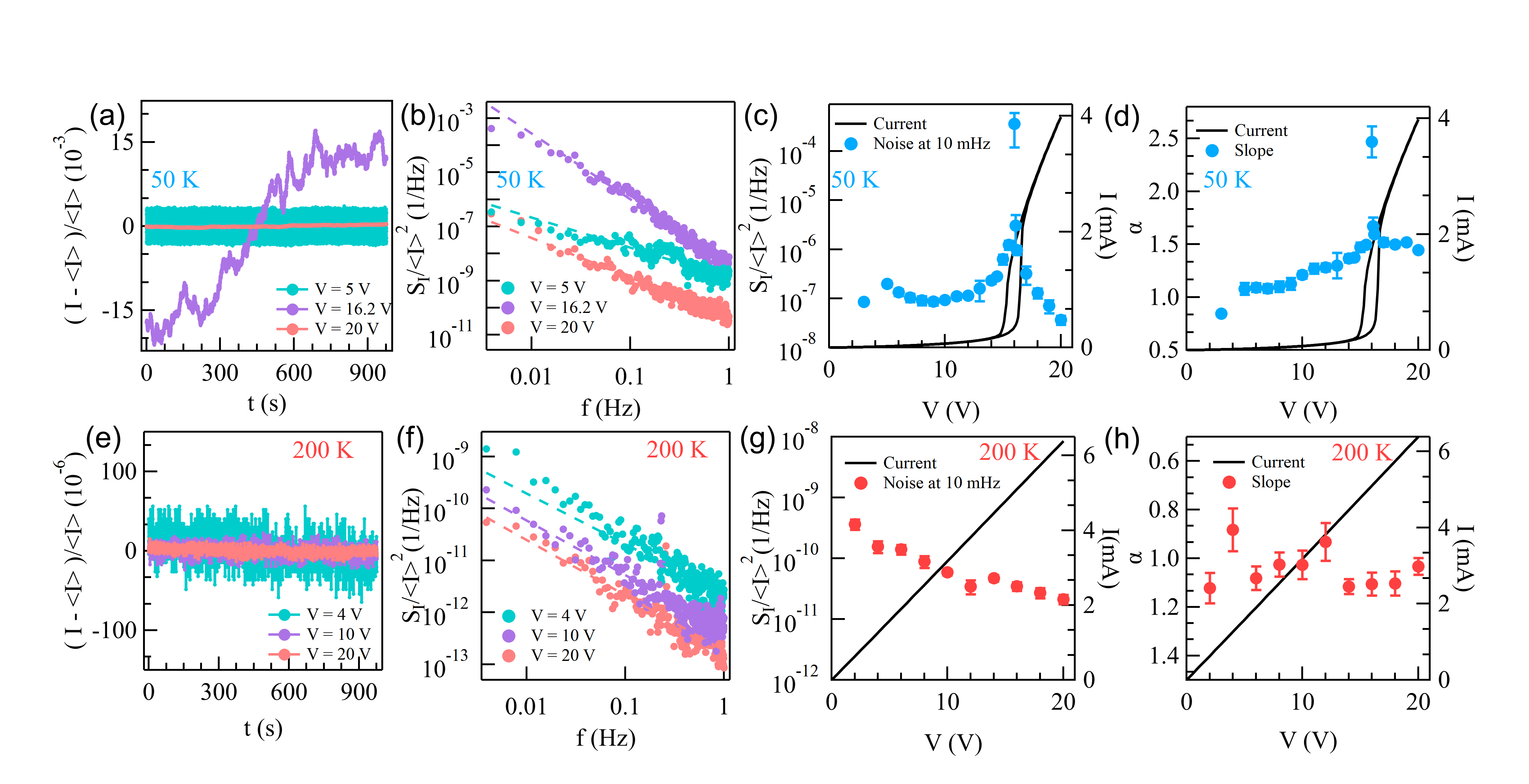}
    \caption{(a-d) Voltage-driven noise behavior at $T = 50$ K: (a) Normalized time series showing the temporal evolution of resistance fluctuations. (b) Normalized power spectral density (PSD) of resistance fluctuations, highlighting $1/f^{\alpha}$ behavior; dashed lines are fits to $1/f^{\alpha}$. (c) Noise magnitude at $f = 10$ mHz as a function of applied voltage. (d) Frequency exponent of the PSD, illustrating changes in the noise spectrum with voltage. (e-h) Voltage-driven noise behavior at $T = 200$ K: (e) Normalized time series showing resistance fluctuations over time. (f) Normalized PSD of resistance fluctuations; dashed lines are fits $1/f^{\alpha}$. (g) Noise magnitude at $f = 10$ mHz as a function of applied voltage. (h) Voltage dependence of the frequency exponent of PSD.}
    \label{fig:voltage_driven_noise}
\end{figure*}

The dynamics of CO breakdown are explored through a detailed analysis of the low-frequency noise during the voltage-driven transition. Fig. \ref{fig:voltage_driven_noise} presents the noise spectroscopy results at two distinct temperatures: $T = 50~\mathrm{K}$, well below $T_\mathrm{CO}$ (Fig. \ref{fig:voltage_driven_noise} (a)-(d)), and $T = 200~\mathrm{K}$ (Fig. \ref{fig:voltage_driven_noise} (e)-(h)), above $T_\mathrm{CO}$.

At $T = 50~\mathrm{K}$, a sharp increase in current is observed at $V \sim 16.5~\mathrm{V}$, as depicted by the black curve in Fig. \ref{fig:voltage_driven_noise}(c) or (d). This nonlinear current–voltage characteristic is often associated with collective transport processes and may signify the breakdown of an ordered charge configuration. The normalized residual current fluctuations in Fig. \ref{fig:voltage_driven_noise}(a) offer insight into the underlying dynamics of the system. At $V = 5~\mathrm{V}$, the time series exhibits minimal fluctuations, suggesting a spatially coherent arrangement of charge carriers, consistent with a charge-localized phase. In contrast, at $V = 16.2~\mathrm{V}$, just prior to the current jump, the fluctuations become pronounced and erratic. This behavior indicates that the external electric field begins to destabilize the charge arrangement, facilitating reconfiguration or the introduction of disorder into the system. Such large fluctuations are compatible with the onset of spatial and temporal rearrangements of charge domains. After the transition, at $V = 20~\mathrm{V}$, these strong fluctuations subside, implying the system has entered a new, more dynamically conductive state in which the previously ordered phase has collapsed.

The corresponding PSD spectra shown in Fig. \ref{fig:voltage_driven_noise}(b) quantitatively captures this evolution. At $V = 5~\mathrm{V}$, the PSD magnitude is low, reflecting the suppressed noise levels associated with a static charge arrangement. As the applied voltage reaches $V = 16.2~\mathrm{V}$, the PSD magnitude increases significantly across all frequencies, and the spectral slope $\alpha$ becomes markedly steeper. This trend is consistent with collective rearrangement or depinning dynamics as the system transitions out of the localized state. The increase in correlation time associated with current fluctuations contribute to the enhanced low-frequency noise. Conversely, at $V = 20~\mathrm{V}$, the PSD magnitude decreases, indicating reduced temporal fluctuation and a more stable conductive phase where the initial charge order has melted.

Further analysis is presented in Fig. \ref{fig:voltage_driven_noise}(c) and (d). In Fig. \ref{fig:voltage_driven_noise}(c), the noise magnitude at $f = 10~\mathrm{mHz}$ remains low at low voltages, but increases sharply as the system approaches $V \sim 16.5~\mathrm{V}$, peaking just before the current discontinuity. The abrupt rise abrupt rise in noise is indicative of critical dynamics often observed during the collapse of an ordered electronic state. In the context of $\beta$-Na$_{0.33}$V$_2$O$_5$, such dynamics are consistent with the field-driven destabilization of the charge-ordered state formed by localized small polarons. It suggests a regime where the applied voltage competes with the intrinsic pinning potential imposed by lattice or impurity interactions. Simultaneously, Fig. \ref{fig:voltage_driven_noise}(d) shows the evolution of the PSD exponent $\alpha$, which is close to 1 at low voltages-typical of $1/f$ noise. As the transition is approached, $\alpha$ increases toward 2, reflecting a shift to noise behavior dominated by slow, diffusive-like dynamics. A $1/f^2$ noise spectrum ($\alpha \approx 2$) is indicative of non-equilibrium dynamics associated with depinning or cooperative motion. Similar behavior has been observed in charge-ordered organic conductors such as $\kappa$-(BETS)$_2$Mn[N(CN)$_2$]$_3$, where a sharp rise in noise and $\alpha \approx 2$ appears below the transition, reflecting slow, correlated fluctuations of polar nanoregions.\cite{Thomas2024}
\begin{figure*}
    \centering
    \includegraphics[trim=10 10 10 10, clip,width=\linewidth]{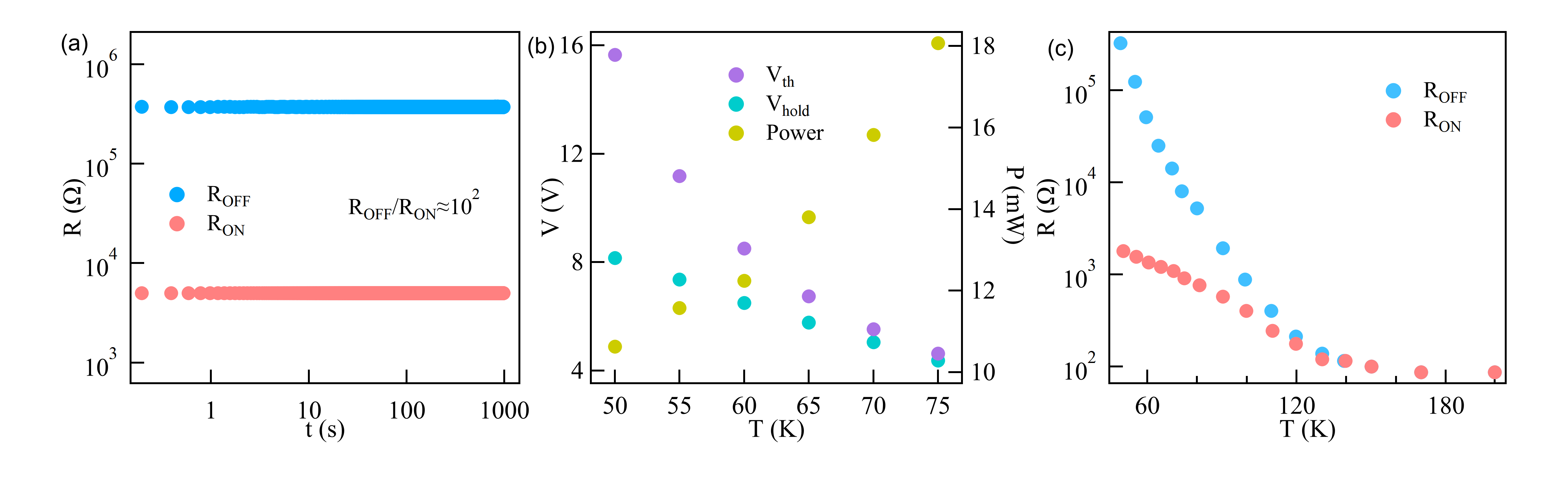} 
    \caption{(a) Resistance retention in the ON and OFF states over time, showing stability up to $10^3$ s. (b) Variation of threshold and hold voltages with temperature. (c) Temperature dependence of $R_{\mathrm{ON}}$ (low-resistance state) and $R_{\mathrm{OFF}}$ (high-resistance state), highlighting the ratio increases with decrease in temperature.}
    \label{fig:resistive_switching}
\end{figure*}

In contrast, at $T = 200~\mathrm{K}$ where the system is above \( T_{CO} \) residual current fluctuations remain largely unchanged across the voltage range, as shown in Fig. \ref{fig:voltage_driven_noise}(e). The PSD spectra in Fig. \ref{fig:voltage_driven_noise}(f) exhibit no appreciable change in either magnitude or slope, and both the noise amplitude and the frequency exponent $\alpha$ plotted in Figs. \ref{fig:voltage_driven_noise}(g) and (h) remain nearly constant throughout. These observations indicate that at $T = 200~\mathrm{K}$, the system remains in a delocalized, homogeneous phase where the applied electric field does not perturb the equilibrium configuration. In contrast, the strong temperature dependence of the noise response at $T = 50~\mathrm{K}$ underscores the critical role of thermal energy of ions and charge carriers in stabilizing or destabilizing the charge arrangement. The noise characteristics at low temperature clearly capture the voltage-driven breakdown of a charge-localized state, while their absence at higher temperature affirms the collapse of this phase well above the ordering temperature. The ability to monitor these noise features offers a powerful diagnostic for probing nonequilibrium transitions and informs design strategies for cryogenic memory technologies, where thermal and electric-field control of conducting states is critical.

The cryogenic memory functionality of $\beta$-Na$_{0.33}$V$_2$O$_5$ relies on the inherently low conductivity of the CO state. As previously discussed, this CO state can be destabilized by an external electric field.
Figure~\ref{fig:resistive_switching}(a) illustrates the volatile resistive switching behavior, where the resistance $R$ is plotted as a function of time $t$. The high-resistance state (HRS), corresponding to the CO phase ($R_{\mathrm{OFF}}$), remains stable for durations exceeding $10^3$ seconds. Upon the application of an electric field, the CO state is suppressed, resulting in a transition to a low-resistance state (LRS, $R_{\mathrm{ON}}$), which is clearly distinguishable from the HRS and similarly stable over the same timescale. The difference in conductivity between the two states leads to a resistance ratio $R_{\mathrm{OFF}} / R_{\mathrm{ON}} \approx 10^2$, offering a pronounced contrast suitable for reliable data retention and readout.

Moreover, the temperature dependence of the threshold voltage $V_{\text{th}}$ shown in Fig.~\ref{fig:resistive_switching}(b), displays an exponential behavior. This trend is consistent with prior observations in related charge-ordered systems,\cite{Yamanouchi1999, Asamitsu1997} where the CO phase becomes increasingly susceptible to electric-field-induced breakdown at lower temperatures.

The rise in $V_{\text{th}}$ with decreasing temperature reflects the increasing stability of the CO polaronic insulating phase, which is disrupted under an applied electric field, leading to a transition into a more conducting phase and the emergence of strongly nonlinear current-voltage behavior.\cite{PARIJA20201166} This behavior is consistent with a field-induced delocalization of polarons and the collapse of the charge-ordered phase. The repeatable and non-destructive nature of the resistive transition, along with the absence of a required forming voltage during the initial application of high bias and the observed threshold-like behavior, suggests that the switching mechanism is not due to dielectric breakdown or filament formation.\cite{Adda2022} Instead, it is more plausibly attributed to the field-induced destabilization of localized polarons into delocalized charge carriers, resulting in the breakdown of the CO state and the onset of band-like conduction.\cite{Bryksin1998, Emin2006} This transition is dominated by electronic effects, as opposed to thermal runaway due to Joule heating. This is supported by examining the power dissipated around the threshold voltage shown in Fig. \ref{fig:resistive_switching} (b). Measurements of the thermal conductivity $\kappa$ of $\beta$-Na$_{0.33}$V$_2$O$_5$ \cite{Markina2004} indicate it increases as the sample cools below $T_{\text{CO}}$, improving the system's ability to disperse heat into the environment through its coupling to the cryostat via the exchange gas within the sample space. Thus, more power would be required to locally heat and trigger the transition if Joule heating were responsible. Instead, we find that the power required to switch the device decreases with cooling.\cite{Lee2008} Both theory and experiments suggest that the threshold switching behavior and the observed field values in the crystals are comparable to those found in other systems consisting of localized charge carriers undergoing hopping conduction, whether driven by polarons or disorder-induced Anderson localization.\cite{Bryksin1998, Ielmini2008} These systems include oxides such as Co$_{1+x}$Cr$_{2-x}$O$_4$, Co$_{1-x}$Li$_x$O$_4$, Pr$_{1-x}$Ca$_x$MnO$_3$, and numerous manganates, La$_{2-x}$Sr$_x$NiO$_4$, transition-metal-oxide glasses, and conjugated polymers, all of which exhibit threshold fields on the order of \( 10^2 \)–\( 10^6 \) V/cm.\cite{Yamanouchi1999, Ielmini2008, Asamitsu1997, Basko2002, Lee2008} Fig. \ref{fig:resistive_switching} (c) shows that the resistance ratio $R_{\mathrm{OFF}} / R_{\mathrm{ON}}$ increases systematically with decreasing temperature, reflecting the progressive stabilization of the CO state in $\beta$-Na$_{0.33}$V$_2$O$_5$. This enhancement in resistive contrast at low temperatures highlights the material’s potential for cryogenic memory applications, where a large and temperature-dependent $R_{\mathrm{OFF}} / R_{\mathrm{ON}}$ ratio is desirable for reliable state discrimination.

In conclusion, our transport and low-frequency noise measurements reveal that charge carriers in $\beta$-Na$_{0.33}$V$_2$O$_5$ behave as small polarons, with a temperature-dependent crossover from thermally activated nearest-neighbor hopping to variable-range hopping below the charge-ordering temperature \( T_{\text{CO}} \). This transition is accompanied by a sharp suppression in noise magnitude, indicating the onset of collective charge dynamics in the ordered state. Structural evidence from temperature-dependent X-ray diffraction confirms the development of a modulated superstructure consistent with ordering of interstitial ions concomitant with CO. Notably, below \( T_{\text{CO}} \), an applied electric field can destabilize the charge-ordered insulating phase, inducing volatile resistive switching into a high-conducting state—an effect marked by abrupt changes in current fluctuations and spectral characteristics. These findings provide direct evidence linking charge order to field-driven electronic switching and offer new insight into the dynamic interplay between localization, correlation, and nonequilibrium transport in low-dimensional oxides. The ability to electrically control such transitions in a reproducible and temperature-dependent manner underscores the potential of charge-ordered vanadium bronzes for cryogenic memory and neuromorphic computing technologies.

\begin{acknowledgments}
Electrical transport measurements at the University at Buffalo were supported by the National Science Foundation grant \#NSF-MRI 1726303. NK and NJ acknowledge support from the College of Arts and Sciences at the University at Buffalo. 
JP and SB acknowledge support from the Center for Reconfigurable Electronic Materials Inspired by Nonlinear Neuron Dynamics (reMIND), an Energy Frontier Research Center funded by the US Department of Energy, Office of Science, Basic Energy Sciences under Award No. DE-SC0023353.
\end{acknowledgments}
\bibliography{Main}
\end{document}